\documentclass[aps,twocolumn,prl,superscriptaddress,letterpaper,nofootinbib]{revtex4-1}
\usepackage{amsmath,amsfonts,amssymb}
\usepackage{mathtools}
\usepackage[pdftex]{graphicx}
\usepackage[usenames, dvipsnames, svgnames, table]{xcolor}
\usepackage[colorlinks=true, citecolor=Blue,
            linkcolor=BrickRed, urlcolor=Brown]{hyperref}
\usepackage{txfonts}
\usepackage{mathrsfs}
\usepackage{bm}
\usepackage{multirow}
\usepackage{braket}
\usepackage{diagbox, eqparbox, hhline}

\newcommand{\be}{\begin{equation}}
\newcommand{\ee}{\end{equation}}
\newcommand{\ben}{\begin{eqnarray}}
\newcommand{\een}{\end{eqnarray}}
\newcommand{\bes}{\begin{subequations}}
\newcommand{\ees}{\end{subequations}}
\newcommand{\bF}{\begin{figure}}
\newcommand{\eF}{\end{figure}}

\newcommand{\avg}[1]{\left\langle{#1}\right\rangle}

\newcommand{\hh}{\hat{H}}

\newcommand{\im}{\mathrm{i}}

\newcommand{\p}{\hat{p}}

\newcommand{\q}{\hat{q}}

\newcommand{\uu}{\hat{U}}

\newcommand{\nth}{N_{\mathrm{th}}}

\begin{document}

\title{Signatures of the quantum nature of gravity in the differential motion of two masses}

\author{Animesh Datta}
\affiliation{Department of Physics, University of Warwick, Coventry CV4 7AL, United Kingdom}

\author{Haixing Miao}
\affiliation{School of Physics and Astronomy \& Institute for Gravitational Wave Astronomy, University of Birmingham, B15 2TT, United Kingdom}

\date{\today}


\begin{abstract}

We show that a signature of the quantum nature of gravity is the quantum mechanical squeezing of the differential motion of two identical masses with respect to their common mode.
This is because the gravitational interaction depends solely on the relative position of the two masses.
In principle, this squeezing is equivalent to quantum entanglement between the masses.
In practice, detecting the squeezing is more feasible than detecting the entanglement.
To that end, we propose an optical interferometric scheme to falsify hypothetical models of gravity.

 \end{abstract}

\maketitle


\emph{Introduction:} 
Beginning with the work of Bronstein in 1936, thought experiments have been a perennial tool in understanding the quantum nature of gravity~\cite{Bronstein1936a,Gorelik2005}.
In his critical comments at the Chapel Hill Conference on \emph{The Role of Gravitation in Physics} in 1957, Feynman concluded that a real problem in a quantum mechanical theory of gravitation is the lack of experimental guidance. 
``In this field since we are not pushed by experiments we must be pulled by imagination.'' he said~\cite{Feynman1957}. 
That era may be about to close. 

The last few years have witnessed increasing numbers of experimental proposals for detecting gravitational effects in quantum systems~\cite{KafriTaylor2013,Anastopoulos2015,Bose2017,Marletto2017,Marletto2018,Balushi2018,Haine2018,Howl2019,Carlesso2019,Carney2019,Krisnanda2020,Miao2020,Clarke2020,Matsumara2020,Liu2021}. 
There have followed several clarifications and analyses~\cite{Hall2018,Anastopoulos2018,Belenchia2018,MarlettoVedral2018,MarlettoVedral2019,Reginatto2019,Christodoulou2019,MarlettoVedral2020,Bose2018,Pedernales2020,Kim2020,Marshman2020,vandeKamp2020}. 
These presage the aftermath of the Page-Geilker experiment in the previous generation~\cite{Page1981,Hawkins1982,Ballentine1982,Page1982}.
While the present proposals are some years from being experimentally implemented and the precise implications of their predictions still debated, it behooves us to seek the simplest and most transparent route for detecting gravitational effects between masses in the quantum regime. 

In this Letter, we show the gravitational interaction between two identical masses leads to quantum mechanical squeezing of their differential mode of motion with respect to its common mode.
Its origin lies in a shift in the frequency of that former mode relative to the latter.
In turn, this motional squeezing is the genesis of the quantum entanglement between two masses interacting gravitationally in the Newtonian limit~\cite{Bose2017,Marletto2017}.
Experimentally, detecting the squeezing is less challenging than detecting the entanglement.
With that in mind, we propose an optical interferometric scheme to test hypothetical models of gravity.
We close by suggesting an electromagnetic version of our scheme to hone the experimental techniques.

\begin{figure}[t]
\includegraphics[width=0.75\columnwidth]{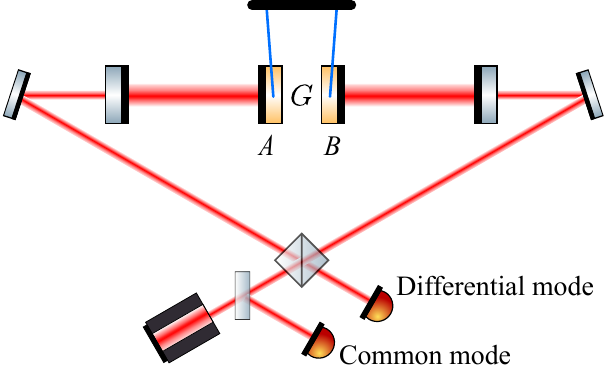}
\caption{Schematic of an optical interferometer for probing quantum signatures of the gravity in the differential motion of two identical masses $A,B$. 
The masses also form the end mirrors of optical cavities.
Quantum signatures of the gravitational interaction can thus be detected optomechanically.
For simplicity, we have omitted the local oscillator for the homodyne measurement of the optical quadratures.}
\label{fig:config}
\end{figure}

\emph{Principle - Motional Squeezing:}
We consider two identical simple harmonic oscillators 
$A$ and $B$ of mass $m$ and frequency $\omega_m$ each, 
separated at equilibrium by a distance $d,$ and located at $\q_A,
\q_B$ with momenta $\hat{p}_A,\hat{p}_B.$ The hamiltonian, \emph{if} the gravitational interaction between the two masses is quantum, in the Newtonian limit is,
\be
\hh= \frac{1}{2m}\left(\hat{p}_A^2+\hat{p}_B^2 \right) + \frac{m\omega_m^2}{2} \left(\q_A^2+\q_B^2\right)   -\frac{G m^2}{|d+\q_A - \q_B|},
\label{eq:master}
\ee
where $[\q_i,\p_j]= \im \hbar \delta_{ij}$ for $i,j = A,B.$
If $d \gg |\q_A - \q_B|,$ expanding the gravitational interaction up to the second order in $\q_A-\q_B,$ neglecting a constant and absorbing a linear term gives
\be
\hh   \equiv   \hh_+ +\hh_- = \frac{\p_+^2}{2m} + \frac{m\omega_m^2}{2}\q_+^2  +  \frac{\p_-^2}{2m} + \frac{m\omega_-^2}{2}\q_-^2 ,
\label{eq:hpm}
\ee
for the common ($+$) and differential ($-$) modes defined as
\be
\p_{\pm} = \frac{\p_A \pm \p_B}{\sqrt{2}}, ~~~ \q_{\pm} = \frac{\q_A \pm \q_B}{\sqrt{2}},
\label{eq:transf}
\ee
and 
\be
\omega_-=\omega_m \sqrt{1-\delta^2},~~ \delta = \frac{2\omega_G}{\omega_m},~~ \omega_G = \sqrt{\frac{Gm}{d^3}}. 
\ee
The gravitational interaction is captured by the characteristic frequency $\omega_G$ that depends on $m$ and $d.$ 
If $d$ is comparable to the size of the masses, $\omega_G$ is independent of the mass. 
Rather, it depends on the material density $\rho$ as~\cite{Miao2020}.
\be
\omega_G = \sqrt{\Lambda G \rho},
\ee
where $\Lambda$ is of the order of unity determined by the geometry of the mass.
For gold, $\omega_G \sim 10^{-3}$ Hz, implying $\delta \ll 1$ even for Hz scale oscillators.

As $[ \hh_+ , \hh_- ] =0$ from Eq.\,\eqref{eq:hpm}, the $\pm$ modes are decoupled.
The gravitational interaction affects only the differential one and manifests itself as a shift in the frequency of that mode relative to the common mode. 
An optical interferometric setup for monitoring the differential mode is presented in Fig.~(\ref{fig:config}).


If the two masses are in thermal states of the simple harmonic hamiltonian at temperature $T$ at the onset of the gravitational interaction, the differential mode is in a Gaussian state with covariance matrix 
\be
{\bm \sigma}^-_0=  \hbar  \nth \left[\begin{array}{cccc}
\dfrac{1}{m\omega_m}  & 0 \\
0  &  m \omega_m
\end{array}\right] ,~~\nth= \frac{1}{2}\coth \left(\frac{\hbar \omega_m}{2 k_B T} \right),
\ee
where $k_B$ is Boltzmann's constant.
As $\delta \ll1$, after time $\tau$ the differential mode evolves under $ \uu_- = \exp(- \im \tau\hh_-/\hbar)$   to
\be
{\bm \sigma}^-_\tau= \hbar  \nth  \left[\begin{array}{cccc}
\dfrac{1+ \delta^2 \sin^2 (\omega_m \tau) }{m\omega_m}	& \dfrac{\delta^2}{2} \sin (2\omega_m \tau)  \\
\dfrac{\delta^2}{2} \sin (2\omega_m \tau) 							& m \omega_m \left(1- \delta^2 \sin^2 (\omega_m \tau) \right)
\end{array}\right].
\label{eq:sigt}
\ee
The fundamental signature of the quantum nature of the gravitational interaction at the Newtonian limit is thus a squeezing of the differential momentum and an equivalent anti-squeezing of the differential position quadrature. 
Additionally, the gravitational interaction generates correlations between the two quadratures. 
All these signatures are proportional to $\delta^2$ and oscillate at double the frequency of the original harmonic potential.
Note that these signatures are a consequence of the frequency shift of the differential mode relative to the common one which remains in its original thermal state for the entire evolution.

Any effort to detect the minuscule quantum mechanical squeezing of the differential motional mode in the laboratory will be marred by decoherence.
Its specific nature will depend on the details of the experimental setup and its regime of operation. 
For instance, under the Caldeira-Leggett model of decoherence for an Ohmic spectral density with an infinite cutoff, the steady-state covariance matrices of the differential and common motions are given by
\be
{\bm \sigma}^-_{\infty}=  \hbar  \nth \left[\begin{array}{cccc}
\dfrac{1}{m\omega_-}  & 0 \\
0  &  m \omega_-
\end{array}\right], ~
{\bm \sigma}^+_{\infty}=  \hbar  \nth \left[\begin{array}{cccc}
\dfrac{1}{m\omega_m}  & 0 \\
0  &  m \omega_m
\end{array}\right],
\label{eq:siginf}
\ee
showing that the only remnant of the quantum aspect of gravity is the altered variances or quantum mechanical squeezing of the differential motional mode - a direct consequence of its shifted frequency due to the gravitational interaction.

Interestingly, this squeezing of the differential motion manifests itself as the entanglement between the two masses upon transformation as per Eq.~(\ref{eq:transf}). 
Mathematically, the entanglement of the two masses $A,B$ is determined by the product of the two smallest eigenvalues of the covariance matrix of the common and differential modes~\cite{WolfEisertPlenio2003}.
Physically, this is identical to the creation of two-mode entanglement by interfering a 
squeezed and an unsqueezed thermal mode on a balanced beam splitter. 
The condition for achieving steady-state quantum entanglement 
is identical (see Eq.~(\ref{eq:ent_cond})), to within a factor of two, to that for observing squeezing.

Observing this squeezing requires, for $\delta \ll 1,$ 
\be
\label{eq:sq}
N_{\rm th} \frac{\omega_-}{\omega_m}\leq \dfrac{1}{2},~\text{or}~\coth \left(\frac{\hbar \omega_m}{2 k_B T} \right)\leq (1-\delta^2)^{-1/2}\approx 1 + \frac{\delta^2}{2}\,.
\ee
This, in turn, requires 
\be
\frac{k_B T}{\hbar \omega_m} \leq \frac{1}{\ln {(4/\delta^2})}\,,
\ee
which may seem impossible to achieve. 
However, as this steady-state behaviour does not depend on 
the mechanical damping or the mechanical quality factor 
$Q_m,$ feedback~\cite{Mancini1998,Cohadon1999} or optomechanical~\cite{Marquardt2007,Wilson-Rae2007} cooling can be invoked to reduce the effective temperature 
to $T/Q_m.$
Consequently, final condition for observing the squeezing becomes 
\be
\label{eq:sqz_cond}
\frac{k_B T}{\hbar \omega_m} \leq \frac{Q_m}{\ln {(4/\delta^2})}\,.
\ee
This is within the reach of current technology using masses of gold or tungsten with $\rho = 19.3 \times 10^3~\text{Kg/m}^3,$ $\Lambda =2,$ 
$ T = 10$~mK, $\omega_m = 2\pi \times 10~\mathrm{Hz}$ , and $Q_m=10^9$. 
However, low frequency mechanical noise is expected to be a challenge. 
To that end,  we now estimate the impacts of gravity-gradient and seismic noise on our proposal.

Gravity-gradient noise due to seismically-induced matter density fluctuation is quite small at $10~\mathrm{Hz}$. 
Even for Advanced LIGO, this is substantially smaller than thermal noise~\cite[Fig.~2]{ALIGO2015}.
A much larger contribution comes from anthropic activity, such as a person walking close to the experiment. 
This was analysed in the context of the gravitational wave detector~\cite{Thorne1999}.
Converting the strain noise to acceleration, and then force noise on a 1 gram mass gives $1.0\times 10^{-16}~\mathrm{N}/\sqrt{\mathrm{Hz}}$ at $10\,\rm Hz$ for a person is moving 1 m away from the experiment. 
This is more than one order of magnitude larger than the thermal Langevin force noise of about $6.0\times 10^{-18}~\mathrm{N}/\sqrt{\mathrm{Hz}}$ at 10\,mK and $10\,\rm Hz.$
This may be mitigated by restraining anthropic activity in the laboratory during data collection. 

To counter seismic noise, the experiment can be seismically isolated to a level close to $10^{-12}\,\rm m/\sqrt{\rm Hz}$ at $10\,\rm Hz$ using a commercial active isolation platform (e.g., Accurion) in a reasonably quiet lab ($10^{-10}\,\rm m/\sqrt{\rm Hz}$ at $10\,\rm Hz$).
Furthermore, the resonant frequency of two masses can be well matched such that a significant part of the seismic noise acts on the common mode of the masses without affecting the differential mode. 
To estimate the latter, a difference $\Delta\omega$ in the frequency of two masses will induce on the differential mode an effective force noise of $2m\omega_- \Delta \omega x_{\mathrm{GND}},$ where $x_{\mathrm{GND}}$ is the seismic motion. The corresponding single-sided power spectral density is $4m^2\omega_-^2 (\Delta \omega)^2 S_{xx},$ where $S_{xx}$ is the single-sided power spectral density of the seismic motion.
Setting this to, say, $1\%$ of the thermal Langevin force noise spectral density $S_{FF} = 4\hbar m \gamma_m \omega_- N_{\rm th},$ where $\gamma_m= \omega_m/Q_m,$ gives
\be
4m^2\omega_-^2 (\Delta \omega)^2 S_{xx} = \frac{1}{100} S_{FF}.
\ee
A seismic noise of $10^{-12}\,\rm m/\sqrt{\rm Hz}$ from active isolation gives $S_{xx} = 10^{-24}\,\rm m^2/Hz.$
A 1 gram, $10\,\rm Hz$ oscillator with $Q_m = 10^9$, then requires $\Delta \omega \sim 5 \,\mu \rm Hz,$ which is about 80 times the mechanical linewidth $\gamma_m= 63\, \rm nHz.$
It is also about 10 times the frequency stability of $0.4 \,\mu \rm Hz$ required to observe optical squeezing by integrating over one month (See next section).
Achieving this $\Delta \omega$ may require experimental innovation and development.

The outstanding challenge is resolving quantum squeezing on the order of $\delta^2 \sim 10^{-9}$ for a $10$ Hz mechanical 
oscillator. We address this next.


\emph{Practice - Optomechanical squeezing:}
A possible route to detecting this minuscule motional squeezing is via optical interferometry that detects the relative frequency shift, as illustrated in Fig.~(\ref{fig:config}).

For low-frequency simple harmonic oscillators, the cavity bandwidth can be much larger than the mechanical frequency - a regime amenable to feedback cooling. 
This regime also allows the cavity mode can be eliminated adiabatically, resulting in the input-output relation for the differential \emph{optical} mode~\cite{Chen2013}
\begin{align}
\label{eq:io}
\hat X_{-}^{\rm out}(t)& = \hat X_{-}^{\rm in}(t)\,, \\
\hat Y_{-}^{\rm out}(t) & = \hat Y_-^{\rm in}(t) +
 (\alpha/\hbar) \,\hat q_-(t)\,,
\end{align}
where $\hat X_-~ (\hat Y_-)$ is the amplitude (phase) quadrature in the two-photon formalism\,\cite{Caves1985a},
$\alpha = 8\sqrt{{\hbar P\omega_0}/(c^2 t_m^2)}$ with
 $P$ the intra-cavity power, $\omega_0$ 
the laser frequency, and $t_m$ 
the amplitude transmissivity of the input mirror of the 
optical cavity. 
The equation of motion of the masses' differential mode is 
\be
\label{eq:eom}
m \ddot {\hat q}_-(t) +m \gamma_m \dot {\hat q}_-(t) + m \omega_-^2  {\hat q}_-(t) = \alpha\, \hat X_{-}^{\rm in}(t) + \hat F_{\rm th}(t)\,,
\ee
where $\hat F_{\rm th}$ is the Langevin force with a spectrum consistent with the Caldeira-Leggett model.

Solving Eqs.~(\ref{eq:io})-(\ref{eq:eom}) in the frequency domain gives
\be
\left[\begin{array}{c}
\hat X_{-}^{\rm out}(\omega)\\
\hat Y_{-}^{\rm out}(\omega) 
\end{array}\right] = 
\left[\begin{array}{cc}
1 & 0 \\
\dfrac{\alpha^2\chi_-(\omega)}{\hbar} & 1 
\end{array}\right] \left[\begin{array}{c}
\hat X_{-}^{\rm in}(\omega)\\
\hat Y_{-}^{\rm in}(\omega) 
\end{array}\right] + \left[\begin{array}{c}
0\\
\dfrac{\alpha\chi_-(\omega)}{\hbar}
\end{array}\right] \hat F_{\rm th}(\omega)\,,
\ee
where $\chi_-(\omega)\equiv [-m(\omega^2+i\,\gamma_m\omega-\omega_-^2)]^{-1}$
is the mechanical susceptibility of the differential motion.
The corresponding covariance matrix for the spectral densities is 
\be
\label{eq:Smat}
\left[\begin{array}{cc}
S_{XX}(\omega) & S_{XY}(\omega)\\
S_{YX}(\omega) & S_{YY}(\omega)
\end{array}\right] 
= \left[ \begin{array}{cc}
1					     & \dfrac{\alpha^2\chi_-^*}{\hbar} \\
\dfrac{\alpha^2\chi_-}{\hbar}  & 1+ \dfrac{\alpha^2|\chi_-|^2}{\hbar^2} \left(\alpha^2 + S_{FF}\right)
\end{array}\right] \,,
\ee
where $S_{FF}$ 
results in the steady-state covariance in  Eq.\,\eqref{eq:siginf} for the differential motion.
The correlation between the amplitude and phase quadrature leads to  
optomechanical squeezing\,\cite{Kimble02, Purdy2013b} of the differential optical mode. 

The common optical mode has a similar covariance matrix with the mechanical susceptibility 
$\chi_+ (\omega)\equiv [-m(\omega^2+i\,\gamma_m\omega-\omega_m^2)]^{-1},$
that is with the mechanical resonance at $\omega_m$ rather than $\omega_-.$ 
The quantum nature of gravity thus manifests itself 
as the difference in the ponderomotive squeezing for the 
common and differential modes of the optical field.

This ponderomotive squeezing can be read out using a homodyne measurement. 
It measures a general quadrature $ \hat X(\theta) = \hat X\cos\theta +\hat Y \sin\theta$
determined by $\theta,$ the phase of the local oscillator.
The corresponding spectral density is $
S(\theta, \omega)= S_{XX}\cos^2\theta + (S_{XY}+S_{YX})\sin\theta\cos\theta +  S_{YY}\sin^2\theta\,, 
$ where we have suppressed the $\omega$ argument on the RHS for brevity.
Its minimum value is
\be
S^{\rm min} (\omega)=\frac{S_{XX} + S_{YY}}{2} - \frac{\sqrt{(S_{XY} + S_{YX})^2 + (S_{XX} - S_{YY})^2}}{2}.
\ee
For the common optical mode, $S_{XY} + S_{YX} =0$ at $\omega = \omega_m,$ 
whereby 
\be
S^{\rm min}_+ (\omega_m)> 1.
\ee
It thus shows no squeezing at the mechanical resonance. 
This is an instance of the blindness of homodyne measurements to complex squeezing that is ponderomotively generated~\cite{Buchmann2016}. 

The differential optical mode, on the contrary, does exhibit squeezing at the $\omega = \omega_m$
due of the gravity-induced frequency shift on the differential motional mode.
Mathematically, as $\alpha \gg 1,$ $S_{YY} \gg S_{XY},S_{YX}$ from Eq.~(\ref{eq:Smat}), and $\alpha^2\gg S_{FF}$
in a regime where the quantum radiation pressure dominates the thermal fluctuation,
\be
\label{eq:Sq}
S^{\rm min}_- (\omega_m) \approx 1 - \left \vert \dfrac{\mathrm{Re}[\chi_-(\omega_m)]}{\chi_-(\omega_m)} \right \vert^2 = \frac{1}{1+Q_m^2 \delta^4 }.
\ee
Estimating this squeezing in practice requires repeating the experiment - that is reaching steady state and measuring $\nu$ times.
This gives an estimate with variance $S^{\rm min}_- (\omega_m) /\nu$ which must equal $1-S^{\rm min}_- (\omega_m) $ to provide an SNR of unity.
Thereby, total time for the experiment is
\be
\label{eq:Sqtime}
t = \frac{\nu}{\gamma_m} = \frac{1}{\omega_m Q_m \delta^4},
\ee
after which time, the differential optical squeezing (in dB) is
\be
\label{eq:SqdB}
{\cal S}_- \equiv - 10 \log_{10} \left[S^{\rm min}_- (\omega_m) \right]
= 10 \log_{10}\left[1 + Q_m^2 \delta^4 \right].
\ee
Both ${\cal S}_-$ and $t$ are solely determined by the mechanical, and not the optical properties of the setup.
For the representative experimental parameters chosen after Eq.~(\ref{eq:sqz_cond}), ${\cal S}_- \approx 9~\text{dB}$ and $t \approx 1~\text{month}.$
This is challenging, but not prohibitive. Indeed, its experimental accessibility is aided by the lack of any thermal fluctuation limitations in a regime dominated by the quantum radiation pressure.
An observation of this squeezing shall provide an experimental witness of the quantum nature of gravity - as we illustrate next for a semiclassical model of gravity.

\emph{Semiclassical gravity:}
A hypothetical model of gravity is a semiclassical one - where a classical spacetime structure is sourced by matter of quantum nature.
The hamiltonian for two identical harmonically trapped masses interacting as per the non-relativistic version of such a model - known as the Schr\"odinger-Newton equation~\cite{Yang2013a} is given by
\ben
\hh^{\mathrm{SN}} &=& \frac{1}{2m} \left(\p_A^2+\p_B^2 \right) + \frac{m\omega_m^2}{2} \left(\q_A^2+\q_B^2 \right)  \\
			&-& C_1(\q_A - \q_B) - C_2 \left( (\q_A - \avg{\q_B})^2 + (\q_B - \avg{\q_A})^2 \right).	\nonumber
\label{eq:sn}
\een
where $C_{1,2}$ are constants depending on $m,d.$ Using Eq.~(\ref{eq:transf}),
\be
\hh^{\mathrm{SN}} = \sum_{i={\pm}} \left( \frac{\p_i^2}{2m} + \frac{m\omega_{SN}^2}{2}\q_i^2 \right), ~~~\omega_{SN} = \omega_m\sqrt{ 1- \frac{2C_2}{m\omega_m^2}}.
\ee
In this case, both the common and differential motional modes undergo identical frequency shifts with respect to $\omega_m$ and quantum squeezing. 
There will be no relative frequency shift between the two modes, both of which after time $\tau$ will have identical covariance matrices given by Eq.~(\ref{eq:sigt}) with $\delta=\sqrt{2C_2/m\omega_m^2}.$
Consequently, there will be no quantum mechanical squeezing in the differential mode relative to the common mode.
The quantum squeezing and correlations are transferred unchanged to the motional quadratures of the two masses, as has been discussed in terms of the transfer of quantum information and uncertainty~\cite{Yang2013a}.
Crucially, there will be no quantum entanglement between the motional states of the masses - as is the case if two identically squeezed thermal states are interfered on a balanced beam splitter.

\emph{Conclusions:}
This Letter has three - the first two holding \emph{if} the gravitational interaction between two masses, as given by the last term in Eq.~(1), is sufficient to test the quantum nature of gravity.

Firstly, the quantum nature of the gravitational interaction between two masses can be probed via the squeezing of their differential motional mode.
This is an alternative to the experimental detection of quantum entanglement generated by gravity~\cite{Bose2017,Marletto2017,Marletto2018,Balushi2018,Carlesso2019,Krisnanda2020,Matsumara2020} which, as is to be expected, is also proportional to $\delta^2.$
 Which one is more fundamental depends on which side of Eq.~(\ref{eq:transf}) one considers more intrinsic.

Secondly, the signature of the quantum nature of gravitational interaction is imprinted in the relative frequency shift of the differential motional mode. 
In principle, such a frequency shift may be inferred from a standard transfer function measurement by introducing a strong classical force to drive the masses. 
In practice, the optical squeezing provides direct evidence of the quantum correlation between the optical fields in the two cavities meditated by gravity~\cite{Miao2020}.
This being absent for some hypothetical models of gravity such as the Schr\"odinger-Newton model allows them to be falsified experimentally.

Thirdly, the proposed experiment in Fig.~(\ref{fig:config}) for detecting the squeezing of the differential optical mode is less demanding than one for detecting entanglement - either in the motional or the optical modes. 
The physical reason lies in the negligibility of the thermal fluctuations in the quantum radiation pressure dominated regime.
An exhaustive mathematical argument is hindered by the variety of experimental setups for detecting gravitationally-induced entanglement.
Consequently, we illustrate this conclusion by considering the entanglement between the optical fields in the two cavities identified with the two masses in Fig.~(\ref{fig:config}).


We begin by defining effective single-mode optical quadratures at $\omega_m$ as~\cite{Miao2020}
$
\hat {\cal X}_k \equiv \hat X_k (\omega_m)\sqrt{\Delta \omega/\pi}, 
 \hat {\cal Y}_k \equiv \hat Y_k (\omega_m)\sqrt{\Delta \omega/\pi}\,,
$
such that $[\hat {\cal X}_k, \hat {\cal Y}^{\dag}_{k'} ] = 2i \delta_{k,k'}$ and $ \hat X_k ~ ( \hat Y_k )$ are the amplitude (phase) quadratures of cavity $ k = A,B.$
The bandwidth $\Delta \omega$ is determined by the integration time and smaller than $\gamma_m$. 
The entanglement between the optical fields in the $A,B$ cavities is determined solely by the off-diagonal block of the optical covariance matrix $\bm \sigma^{\rm opt}$ for 
$\{\hat {\cal X}_A, \hat {\cal Y}_A, \hat {\cal X}_B, \hat {\cal Y}_B\}$ quadratures, that is
\be
{\bm \sigma}_{AB}^{\rm opt} =
 \frac{\alpha^2}{2\hbar} \left[\begin{array}{cc} 
 0& \chi^*_+(\omega_m)  - \chi^*_-(\omega_m) \\ 
\chi_+(\omega_m)  - \chi_-(\omega_m)  & \dfrac{\alpha^2}{\hbar}(|\chi_+(\omega_m)|^2 - |\chi_-(\omega_m)|^2)
\end{array}\right]\,. 
\ee
Evidently, it is the difference in the mechanical susceptibilities of 
the common and differential motion that generates the entanglement\footnote{The Schr\"odinger-Newton model predicts $\chi_+(\omega)  = \chi_-(\omega),$ whereby ${\bm \sigma}_{AB}^{\rm opt} = {\bm 0}_{2\times 2},$ and therefore no entanglement. }.
The condition for entanglement between the optical modes, as per the logarithmic negativity, gives
\be
\label{eq:opt_ent_cond}
\coth \left(\frac{\hbar \omega_m }{2 k_B T} \right) \leq Q_m \delta^2,~\text{or}~\frac{k_B T}{\hbar \omega_m} \leq  \left[ \ln\left(\frac{Q_m \delta^2+1}{Q_m \delta^2-1}\right) \right]^{-1}.
\ee
Once again, this is solely determined by the mechanical, and not the optical properties of the setup.
More importantly, this condition on the temperature is `exponentially' more demanding than Eq.~(\ref{eq:sqz_cond}) as well as Eq.~(\ref{eq:Sqtime}), substantiating our second conclusion.
The experimental requirements are summarised in Table~(\ref{SSTable}).
\begin{table}[t!]
\begin{center}
\renewcommand{\arraystretch}{2.0}
\begin{tabular}{ | >{\centering}p{2.0cm} | >{\centering}p{2.0cm} | >{\centering}p{2.0cm} |  }
\hline
 \diagbox[width=2.0cm, height=1.0cm]{Probe}{\raisebox{1.5ex}{Signature}} &	Squeezing			&	  Entanglement 	\tabularnewline \hline
Motional 		& Eq.~(\ref{eq:sqz_cond})  &    Eq.~(\ref{eq:mot_ent_cond})  			\tabularnewline \hline
Optical	        & Eq.~(\ref{eq:Sqtime})	 &  Eq.~(\ref{eq:opt_ent_cond})			 \tabularnewline \hline
\end{tabular}
\caption{Experimental requirements necessary to detect signatures of the quantum nature of gravity.
Squeezing corresponds to that of the differential mode while entanglement is between the two masses. 
Only the optical squeezing imposes no restrictions on the temperature.
Thus, in principle, the experiment can be performed at room temperature if the radiation pressure limited regime can be achieved with low frequency, high quality factor mechanical oscillators.}
\label{SSTable}
\end{center}
\end{table}

Finally, the scheme in Fig.~(\ref{fig:config}) accumulates the quantum information about $\delta$  in the squeezing of the optical differential mode quantum mechanically, 
while the conditional squeezing in Ref.~\cite{Miao2020} combines that same quantum information from the readouts of two optomechanical cavities classically.
The coherence of our scheme leads to a well-known improvement in the detection of $\delta$ by a factor of $\sqrt 2.$
Given the quartic dependences in Eqs.~(\ref{eq:Sqtime}), (\ref{eq:SqdB}), our scheme offers a very consequential four-fold improvement over Ref.~\cite{Miao2020}
\footnote{Eqs.~(2), (3) in Ref.~\cite{Miao2020} are incorrect by a factor of 4.}.

Nevertheless, the experiment proposed here is an exacting one. 
We therefore suggest, as a point of embarkation, an endeavour to detect the relative frequency shift between the common and differential modes due to an electromagentic interaction between the masses. Given its strength, larger values of $\delta$ would be possible with significantly higher $\omega_m.$
This will enable the identification - and suppression, of various noise sources as the electromagnetic strength and $\omega_m$ are progressively reduced.
A course to experimentally detecting signatures of the quantum nature of gravity may thus be charted.



\emph{Appendix - Quantum entanglement between the two masses: }
The steady-state covariance matrix of the masses $A$, $B$, that is for $\{\hat {q}_A, \hat {p}_A, \hat {q}_B, \hat {p}_B\}$ quadratures, is
\be\label{eq:cov}
{\bm \sigma} = 
\left[\begin{array}{cc}
{\bm \sigma}_A & {\bm \sigma}_{AB} \\
{\bm \sigma}_{AB}  & {\bm \sigma}_B 
\end{array}
\right] = {\bf M} \left[\begin{array}{cc}
{\bm \sigma}^{+}_{\infty} & {\bm 0}_{2\times 2} \\
 {\bm 0}_{2\times 2} & {\bm \sigma}^{-}_{\infty} 
\end{array}
\right] {\bf M} \,, 
\ee
where the transformation matrix $\bf M$ is given by 
\be
{\bf M} =\left[\begin{array}{cccc}
\sqrt{\frac{m\omega_m}{\hbar}} & 0 & \sqrt{\frac{m\omega_m}{\hbar}} & 0 \\
0 & \sqrt{\frac{1}{\hbar m\omega_m}} & 0 & \sqrt{\frac{1}{\hbar m\omega_m}} \\
\sqrt{\frac{m\omega_m}{\hbar}}  & 0 & -\sqrt{\frac{m\omega_m}{\hbar}} & 0 \\
0 & \sqrt{\frac{1}{\hbar m\omega_m}} & 0 & -\sqrt{\frac{1}{\hbar m\omega_m}}
\end{array}\right]\,. 
\ee
The quantum entanglement of this bipartite system, as per the logarithmic negativity~\cite{Adesso2007}, is 
$
{\cal E}_N = {\rm max}\left\{-(1/2) \log_2 \left[\left(\Sigma-\sqrt{\Sigma^2-4{\det \bm\sigma}}\right)/2\right] ,\, 0\right\}\,, 
$
where $\Sigma \equiv \det {\bm \sigma}_A +\det {\bm \sigma}_B -2\,\det {\bm \sigma}_{AB}$. 
Using Eq.\,\eqref{eq:siginf},
${\cal E}_N = {\rm max}\left\{-\log_2\left[2N_{\rm th} (1-\delta^2)^{1/4}\right] ,\, 0\right\}.$ 
Achieving steady-state quantum entanglement thus requires
\be
\label{eq:ent_cond}
\coth \left(\frac{\hbar \omega_m}{2 k_B T} \right)\leq (1-\delta^2)^{-1/4}\approx 1 + \frac{\delta^2}{4}\,, 
\ee
which matches Eq.~\eqref{eq:sq} to within a factor of 2. Invoking feedback or optomechanical cooling leads to 
\be
\label{eq:mot_ent_cond}
\frac{k_B T}{\hbar \omega_m} \leq \frac{Q_m}{\ln {(8/\delta^2})}\,.
\ee


\textit{Acknowledgements:}
We thank H. Grote for helpful comments and the UK STFC `Quantum Technologies for Fundamental Physics' programme (ST/T006404/1) for support.
H.M acknowledges the support from the  Birmingham  Institute  for  Gravitational  Wave  Astronomy.  H.M has also been supported by UK STFC Ernest Rutherford Fellowship (Grant No. ST/M005844/11).
We also thank the anonymous referees for comments that helped us improve this Letter.


\bibliography{references}

\begin{thebibliography}{49}%
\makeatletter
\providecommand \@ifxundefined [1]{%
 \@ifx{#1\undefined}
}%
\providecommand \@ifnum [1]{%
 \ifnum #1\expandafter \@firstoftwo
 \else \expandafter \@secondoftwo
 \fi
}%
\providecommand \@ifx [1]{%
 \ifx #1\expandafter \@firstoftwo
 \else \expandafter \@secondoftwo
 \fi
}%
\providecommand \natexlab [1]{#1}%
\providecommand \enquote  [1]{``#1''}%
\providecommand \bibnamefont  [1]{#1}%
\providecommand \bibfnamefont [1]{#1}%
\providecommand \citenamefont [1]{#1}%
\providecommand \href@noop [0]{\@secondoftwo}%
\providecommand \href [0]{\begingroup \@sanitize@url \@href}%
\providecommand \@href[1]{\@@startlink{#1}\@@href}%
\providecommand \@@href[1]{\endgroup#1\@@endlink}%
\providecommand \@sanitize@url [0]{\catcode `\\12\catcode `\$12\catcode
  `\&12\catcode `\#12\catcode `\^12\catcode `\_12\catcode `\%12\relax}%
\providecommand \@@startlink[1]{}%
\providecommand \@@endlink[0]{}%
\providecommand \url  [0]{\begingroup\@sanitize@url \@url }%
\providecommand \@url [1]{\endgroup\@href {#1}{\urlprefix }}%
\providecommand \urlprefix  [0]{URL }%
\providecommand \Eprint [0]{\href }%
\providecommand \doibase [0]{http://dx.doi.org/}%
\providecommand \selectlanguage [0]{\@gobble}%
\providecommand \bibinfo  [0]{\@secondoftwo}%
\providecommand \bibfield  [0]{\@secondoftwo}%
\providecommand \translation [1]{[#1]}%
\providecommand \BibitemOpen [0]{}%
\providecommand \bibitemStop [0]{}%
\providecommand \bibitemNoStop [0]{.\EOS\space}%
\providecommand \EOS [0]{\spacefactor3000\relax}%
\providecommand \BibitemShut  [1]{\csname bibitem#1\endcsname}%
\let\auto@bib@innerbib\@empty
\bibitem [{\citenamefont {Bronstein}(2011)}]{Bronstein1936a}%
  \BibitemOpen
  \bibfield  {author} {\bibinfo {author} {\bibfnamefont {M.}~\bibnamefont
  {Bronstein}},\ }\href {\doibase 10.1007/s10714-011-1285-4} {\bibfield
  {journal} {\bibinfo  {journal} {General Relativity and Gravitation}\ }\textbf
  {\bibinfo {volume} {44}},\ \bibinfo {pages} {267} (\bibinfo {year}
  {2011})}\BibitemShut {NoStop}%
\bibitem [{\citenamefont {Gorelik}(2005)}]{Gorelik2005}%
  \BibitemOpen
  \bibfield  {author} {\bibinfo {author} {\bibfnamefont {G.~E.}\ \bibnamefont
  {Gorelik}},\ }\href {\doibase 10.1070/pu2005v048n10abeh005820} {\bibfield
  {journal} {\bibinfo  {journal} {Physics-Uspekhi}\ }\textbf {\bibinfo {volume}
  {48}},\ \bibinfo {pages} {1039} (\bibinfo {year} {2005})}\BibitemShut
  {NoStop}%
\bibitem [{\citenamefont {{R. P. Feynman}}(1957)}]{Feynman1957}%
  \BibitemOpen
  \bibfield  {author} {\bibinfo {author} {\bibnamefont {{R. P. Feynman}}},\
  }in\ \href@noop {} {\emph {\bibinfo {booktitle} {Chapel Hill Conference
  Proceedings}}}\ (\bibinfo {year} {1957})\BibitemShut {NoStop}%
\bibitem [{\citenamefont {Kafri}\ and\ \citenamefont
  {Taylor}(2013)}]{KafriTaylor2013}%
  \BibitemOpen
  \bibfield  {author} {\bibinfo {author} {\bibfnamefont {D.}~\bibnamefont
  {Kafri}}\ and\ \bibinfo {author} {\bibfnamefont {J.~M.}\ \bibnamefont
  {Taylor}},\ }\href {https://arxiv.org/abs/1311.4558} {\bibfield  {journal}
  {\bibinfo  {journal} {arXiv:1311.4558}\ } (\bibinfo {year}
  {2013})}\BibitemShut {NoStop}%
\bibitem [{\citenamefont {Anastopoulos}\ and\ \citenamefont
  {Hu}(2015)}]{Anastopoulos2015}%
  \BibitemOpen
  \bibfield  {author} {\bibinfo {author} {\bibfnamefont {C.}~\bibnamefont
  {Anastopoulos}}\ and\ \bibinfo {author} {\bibfnamefont {B.~L.}\ \bibnamefont
  {Hu}},\ }\href {\doibase 10.1088/0264-9381/32/16/165022} {\bibfield
  {journal} {\bibinfo  {journal} {Classical and Quantum Gravity}\ }\textbf
  {\bibinfo {volume} {32}},\ \bibinfo {pages} {165022} (\bibinfo {year}
  {2015})}\BibitemShut {NoStop}%
\bibitem [{\citenamefont {Bose}\ \emph {et~al.}(2017)\citenamefont {Bose},
  \citenamefont {Mazumdar}, \citenamefont {Morley}, \citenamefont {Ulbricht},
  \citenamefont {Toro{\v{s}}}, \citenamefont {Paternostro}, \citenamefont
  {Geraci}, \citenamefont {Barker}, \citenamefont {Kim},\ and\ \citenamefont
  {Milburn}}]{Bose2017}%
  \BibitemOpen
  \bibfield  {author} {\bibinfo {author} {\bibfnamefont {S.}~\bibnamefont
  {Bose}}, \bibinfo {author} {\bibfnamefont {A.}~\bibnamefont {Mazumdar}},
  \bibinfo {author} {\bibfnamefont {G.~W.}\ \bibnamefont {Morley}}, \bibinfo
  {author} {\bibfnamefont {H.}~\bibnamefont {Ulbricht}}, \bibinfo {author}
  {\bibfnamefont {M.}~\bibnamefont {Toro{\v{s}}}}, \bibinfo {author}
  {\bibfnamefont {M.}~\bibnamefont {Paternostro}}, \bibinfo {author}
  {\bibfnamefont {A.~A.}\ \bibnamefont {Geraci}}, \bibinfo {author}
  {\bibfnamefont {P.~F.}\ \bibnamefont {Barker}}, \bibinfo {author}
  {\bibfnamefont {M.~S.}\ \bibnamefont {Kim}}, \ and\ \bibinfo {author}
  {\bibfnamefont {G.}~\bibnamefont {Milburn}},\ }\href {\doibase
  10.1103/PhysRevLett.119.240401} {\bibfield  {journal} {\bibinfo  {journal}
  {Phys. Rev. Lett.}\ }\textbf {\bibinfo {volume} {119}},\ \bibinfo {pages}
  {240401} (\bibinfo {year} {2017})}\BibitemShut {NoStop}%
\bibitem [{\citenamefont {Marletto}\ and\ \citenamefont
  {Vedral}(2017)}]{Marletto2017}%
  \BibitemOpen
  \bibfield  {author} {\bibinfo {author} {\bibfnamefont {C.}~\bibnamefont
  {Marletto}}\ and\ \bibinfo {author} {\bibfnamefont {V.}~\bibnamefont
  {Vedral}},\ }\href {\doibase 10.1103/PhysRevLett.119.240402} {\bibfield
  {journal} {\bibinfo  {journal} {Phys. Rev. Lett.}\ }\textbf {\bibinfo
  {volume} {119}},\ \bibinfo {pages} {240402} (\bibinfo {year}
  {2017})}\BibitemShut {NoStop}%
\bibitem [{\citenamefont {Marletto}\ \emph {et~al.}(2018)\citenamefont
  {Marletto}, \citenamefont {Vedral},\ and\ \citenamefont
  {Deutsch}}]{Marletto2018}%
  \BibitemOpen
  \bibfield  {author} {\bibinfo {author} {\bibfnamefont {C.}~\bibnamefont
  {Marletto}}, \bibinfo {author} {\bibfnamefont {V.}~\bibnamefont {Vedral}}, \
  and\ \bibinfo {author} {\bibfnamefont {D.}~\bibnamefont {Deutsch}},\ }\href
  {\doibase 10.1088/1367-2630/aad5d8} {\bibfield  {journal} {\bibinfo
  {journal} {New Journal of Physics}\ }\textbf {\bibinfo {volume} {20}},\
  \bibinfo {pages} {083011} (\bibinfo {year} {2018})}\BibitemShut {NoStop}%
\bibitem [{\citenamefont {{Al Balushi}}\ \emph {et~al.}(2018)\citenamefont {{Al
  Balushi}}, \citenamefont {Cong},\ and\ \citenamefont {Mann}}]{Balushi2018}%
  \BibitemOpen
  \bibfield  {author} {\bibinfo {author} {\bibfnamefont {A.}~\bibnamefont {{Al
  Balushi}}}, \bibinfo {author} {\bibfnamefont {W.}~\bibnamefont {Cong}}, \
  and\ \bibinfo {author} {\bibfnamefont {R.~B.}\ \bibnamefont {Mann}},\ }\href
  {\doibase 10.1103/PhysRevA.98.043811} {\bibfield  {journal} {\bibinfo
  {journal} {Phys. Rev. A}\ }\textbf {\bibinfo {volume} {98}},\ \bibinfo
  {pages} {043811} (\bibinfo {year} {2018})}\BibitemShut {NoStop}%
\bibitem [{\citenamefont {Haine}(2021)}]{Haine2018}%
  \BibitemOpen
  \bibfield  {author} {\bibinfo {author} {\bibfnamefont {S.~A.}\ \bibnamefont
  {Haine}},\ }\href {\doibase 10.1088/1367-2630/abd97d} {\bibfield  {journal}
  {\bibinfo  {journal} {New Journal of Physics}\ }\textbf {\bibinfo {volume}
  {23}},\ \bibinfo {pages} {033020} (\bibinfo {year} {2021})}\BibitemShut
  {NoStop}%
\bibitem [{\citenamefont {Howl}\ \emph {et~al.}(2019)\citenamefont {Howl},
  \citenamefont {Penrose},\ and\ \citenamefont {Fuentes}}]{Howl2019}%
  \BibitemOpen
  \bibfield  {author} {\bibinfo {author} {\bibfnamefont {R.}~\bibnamefont
  {Howl}}, \bibinfo {author} {\bibfnamefont {R.}~\bibnamefont {Penrose}}, \
  and\ \bibinfo {author} {\bibfnamefont {I.}~\bibnamefont {Fuentes}},\ }\href
  {\doibase 10.1088/1367-2630/ab104a} {\bibfield  {journal} {\bibinfo
  {journal} {New Journal of Physics}\ }\textbf {\bibinfo {volume} {21}},\
  \bibinfo {pages} {043047} (\bibinfo {year} {2019})}\BibitemShut {NoStop}%
\bibitem [{\citenamefont {Carlesso}\ \emph {et~al.}(2019)\citenamefont
  {Carlesso}, \citenamefont {Bassi}, \citenamefont {Paternostro},\ and\
  \citenamefont {Ulbricht}}]{Carlesso2019}%
  \BibitemOpen
  \bibfield  {author} {\bibinfo {author} {\bibfnamefont {M.}~\bibnamefont
  {Carlesso}}, \bibinfo {author} {\bibfnamefont {A.}~\bibnamefont {Bassi}},
  \bibinfo {author} {\bibfnamefont {M.}~\bibnamefont {Paternostro}}, \ and\
  \bibinfo {author} {\bibfnamefont {H.}~\bibnamefont {Ulbricht}},\ }\href
  {\doibase 10.1088/1367-2630/ab41c1} {\bibfield  {journal} {\bibinfo
  {journal} {New Journal of Physics}\ }\textbf {\bibinfo {volume} {21}},\
  \bibinfo {pages} {093052} (\bibinfo {year} {2019})}\BibitemShut {NoStop}%
\bibitem [{\citenamefont {Carney}\ \emph {et~al.}(2019)\citenamefont {Carney},
  \citenamefont {Stamp},\ and\ \citenamefont {Taylor}}]{Carney2019}%
  \BibitemOpen
  \bibfield  {author} {\bibinfo {author} {\bibfnamefont {D.}~\bibnamefont
  {Carney}}, \bibinfo {author} {\bibfnamefont {P.~C.~E.}\ \bibnamefont
  {Stamp}}, \ and\ \bibinfo {author} {\bibfnamefont {J.~M.}\ \bibnamefont
  {Taylor}},\ }\href {\doibase 10.1088/1361-6382/aaf9ca} {\bibfield  {journal}
  {\bibinfo  {journal} {Class. Quant Grav.}\ }\textbf {\bibinfo {volume}
  {36}},\ \bibinfo {pages} {034001} (\bibinfo {year} {2019})}\BibitemShut
  {NoStop}%
\bibitem [{\citenamefont {Krisnanda}\ \emph {et~al.}(2020)\citenamefont
  {Krisnanda}, \citenamefont {Tham}, \citenamefont {Paternostro},\ and\
  \citenamefont {Paterek}}]{Krisnanda2020}%
  \BibitemOpen
  \bibfield  {author} {\bibinfo {author} {\bibfnamefont {T.}~\bibnamefont
  {Krisnanda}}, \bibinfo {author} {\bibfnamefont {G.~Y.}\ \bibnamefont {Tham}},
  \bibinfo {author} {\bibfnamefont {M.}~\bibnamefont {Paternostro}}, \ and\
  \bibinfo {author} {\bibfnamefont {T.}~\bibnamefont {Paterek}},\ }\href
  {https://doi.org/10.1038/s41534-020-0243-y} {\bibfield  {journal} {\bibinfo
  {journal} {npj Quantum Information}\ }\textbf {\bibinfo {volume} {6}},\
  \bibinfo {pages} {12} (\bibinfo {year} {2020})}\BibitemShut {NoStop}%
\bibitem [{\citenamefont {Miao}\ \emph {et~al.}(2020)\citenamefont {Miao},
  \citenamefont {Martynov}, \citenamefont {Yang},\ and\ \citenamefont
  {Datta}}]{Miao2020}%
  \BibitemOpen
  \bibfield  {author} {\bibinfo {author} {\bibfnamefont {H.}~\bibnamefont
  {Miao}}, \bibinfo {author} {\bibfnamefont {D.}~\bibnamefont {Martynov}},
  \bibinfo {author} {\bibfnamefont {H.}~\bibnamefont {Yang}}, \ and\ \bibinfo
  {author} {\bibfnamefont {A.}~\bibnamefont {Datta}},\ }\href {\doibase
  10.1103/PhysRevA.101.063804} {\bibfield  {journal} {\bibinfo  {journal}
  {Phys. Rev. A}\ }\textbf {\bibinfo {volume} {101}},\ \bibinfo {pages}
  {063804} (\bibinfo {year} {2020})}\BibitemShut {NoStop}%
\bibitem [{\citenamefont {Clarke}\ \emph {et~al.}(2020)\citenamefont {Clarke},
  \citenamefont {Sahium}, \citenamefont {Khosla}, \citenamefont {Pikovski},
  \citenamefont {Kim},\ and\ \citenamefont {Vanner}}]{Clarke2020}%
  \BibitemOpen
  \bibfield  {author} {\bibinfo {author} {\bibfnamefont {J.}~\bibnamefont
  {Clarke}}, \bibinfo {author} {\bibfnamefont {P.}~\bibnamefont {Sahium}},
  \bibinfo {author} {\bibfnamefont {K.~E.}\ \bibnamefont {Khosla}}, \bibinfo
  {author} {\bibfnamefont {I.}~\bibnamefont {Pikovski}}, \bibinfo {author}
  {\bibfnamefont {M.~S.}\ \bibnamefont {Kim}}, \ and\ \bibinfo {author}
  {\bibfnamefont {M.~R.}\ \bibnamefont {Vanner}},\ }\href {\doibase
  10.1088/1367-2630/ab7ddd} {\bibfield  {journal} {\bibinfo  {journal} {New
  Journal of Physics}\ }\textbf {\bibinfo {volume} {22}},\ \bibinfo {pages}
  {063001} (\bibinfo {year} {2020})}\BibitemShut {NoStop}%
\bibitem [{\citenamefont {Matsumura}\ and\ \citenamefont
  {Yamamoto}(2020)}]{Matsumara2020}%
  \BibitemOpen
  \bibfield  {author} {\bibinfo {author} {\bibfnamefont {A.}~\bibnamefont
  {Matsumura}}\ and\ \bibinfo {author} {\bibfnamefont {K.}~\bibnamefont
  {Yamamoto}},\ }\href {\doibase 10.1103/PhysRevD.102.106021} {\bibfield
  {journal} {\bibinfo  {journal} {Phys. Rev. D}\ }\textbf {\bibinfo {volume}
  {102}},\ \bibinfo {pages} {106021} (\bibinfo {year} {2020})}\BibitemShut
  {NoStop}%
\bibitem [{\citenamefont {Liu}\ \emph {et~al.}(2021)\citenamefont {Liu},
  \citenamefont {Mummery}, \citenamefont {Zhou},\ and\ \citenamefont
  {Sillanp\"a\"a}}]{Liu2021}%
  \BibitemOpen
  \bibfield  {author} {\bibinfo {author} {\bibfnamefont {Y.}~\bibnamefont
  {Liu}}, \bibinfo {author} {\bibfnamefont {J.}~\bibnamefont {Mummery}},
  \bibinfo {author} {\bibfnamefont {J.}~\bibnamefont {Zhou}}, \ and\ \bibinfo
  {author} {\bibfnamefont {M.~A.}\ \bibnamefont {Sillanp\"a\"a}},\ }\href
  {\doibase 10.1103/PhysRevApplied.15.034004} {\bibfield  {journal} {\bibinfo
  {journal} {Phys. Rev. Applied}\ }\textbf {\bibinfo {volume} {15}},\ \bibinfo
  {pages} {034004} (\bibinfo {year} {2021})}\BibitemShut {NoStop}%
\bibitem [{\citenamefont {Hall}\ and\ \citenamefont
  {Reginatto}(2018)}]{Hall2018}%
  \BibitemOpen
  \bibfield  {author} {\bibinfo {author} {\bibfnamefont {M.~J.~W.}\
  \bibnamefont {Hall}}\ and\ \bibinfo {author} {\bibfnamefont {M.}~\bibnamefont
  {Reginatto}},\ }\href {\doibase 10.1088/1751-8121/aaa734} {\bibfield
  {journal} {\bibinfo  {journal} {Journal of Physics A: Mathematical and
  Theoretical}\ }\textbf {\bibinfo {volume} {51}},\ \bibinfo {pages} {085303}
  (\bibinfo {year} {2018})}\BibitemShut {NoStop}%
\bibitem [{\citenamefont {Anastopoulos}\ and\ \citenamefont
  {Hu}(2018)}]{Anastopoulos2018}%
  \BibitemOpen
  \bibfield  {author} {\bibinfo {author} {\bibfnamefont {C.}~\bibnamefont
  {Anastopoulos}}\ and\ \bibinfo {author} {\bibfnamefont {B.}~\bibnamefont
  {Hu}},\ }\href {https://arxiv.org/abs/1804.11315} {\bibfield  {journal}
  {\bibinfo  {journal} {arXiv:1804.11315}\ } (\bibinfo {year}
  {2018})}\BibitemShut {NoStop}%
\bibitem [{\citenamefont {Belenchia}\ \emph {et~al.}(2018)\citenamefont
  {Belenchia}, \citenamefont {Wald}, \citenamefont {Giacomini}, \citenamefont
  {Castro-Ruiz}, \citenamefont {Brukner},\ and\ \citenamefont
  {Aspelmeyer}}]{Belenchia2018}%
  \BibitemOpen
  \bibfield  {author} {\bibinfo {author} {\bibfnamefont {A.}~\bibnamefont
  {Belenchia}}, \bibinfo {author} {\bibfnamefont {R.~M.}\ \bibnamefont {Wald}},
  \bibinfo {author} {\bibfnamefont {F.}~\bibnamefont {Giacomini}}, \bibinfo
  {author} {\bibfnamefont {E.}~\bibnamefont {Castro-Ruiz}}, \bibinfo {author}
  {\bibfnamefont {{\v{C}}.}~\bibnamefont {Brukner}}, \ and\ \bibinfo {author}
  {\bibfnamefont {M.}~\bibnamefont {Aspelmeyer}},\ }\href
  {https://doi.org/10.1103/physrevd.98.126009} {\bibfield  {journal} {\bibinfo
  {journal} {Physical Review D}\ }\textbf {\bibinfo {volume} {98}} (\bibinfo
  {year} {2018})}\BibitemShut {NoStop}%
\bibitem [{\citenamefont {Marletto}\ and\ \citenamefont
  {Vedral}(2018)}]{MarlettoVedral2018}%
  \BibitemOpen
  \bibfield  {author} {\bibinfo {author} {\bibfnamefont {C.}~\bibnamefont
  {Marletto}}\ and\ \bibinfo {author} {\bibfnamefont {V.}~\bibnamefont
  {Vedral}},\ }\href {\doibase 10.1103/PhysRevD.98.046001} {\bibfield
  {journal} {\bibinfo  {journal} {Phys. Rev. D}\ }\textbf {\bibinfo {volume}
  {98}},\ \bibinfo {pages} {046001} (\bibinfo {year} {2018})}\BibitemShut
  {NoStop}%
\bibitem [{\citenamefont {Marletto}\ and\ \citenamefont
  {Vedral}(2019)}]{MarlettoVedral2019}%
  \BibitemOpen
  \bibfield  {author} {\bibinfo {author} {\bibfnamefont {C.}~\bibnamefont
  {Marletto}}\ and\ \bibinfo {author} {\bibfnamefont {V.}~\bibnamefont
  {Vedral}},\ }\href {https://arxiv.org/abs/1907.08994} {\bibfield  {journal}
  {\bibinfo  {journal} {arXiv:1907.08994}\ } (\bibinfo {year}
  {2019})}\BibitemShut {NoStop}%
\bibitem [{\citenamefont {Reginatto}\ and\ \citenamefont
  {Hall}(2019)}]{Reginatto2019}%
  \BibitemOpen
  \bibfield  {author} {\bibinfo {author} {\bibfnamefont {M.}~\bibnamefont
  {Reginatto}}\ and\ \bibinfo {author} {\bibfnamefont {M.~J.~W.}\ \bibnamefont
  {Hall}},\ }\href {\doibase 10.1088/1742-6596/1275/1/012039} {\bibfield
  {journal} {\bibinfo  {journal} {Journal of Physics: Conference Series}\
  }\textbf {\bibinfo {volume} {1275}},\ \bibinfo {pages} {012039} (\bibinfo
  {year} {2019})}\BibitemShut {NoStop}%
\bibitem [{\citenamefont {Christodoulou}\ and\ \citenamefont
  {Rovelli}(2019)}]{Christodoulou2019}%
  \BibitemOpen
  \bibfield  {author} {\bibinfo {author} {\bibfnamefont {M.}~\bibnamefont
  {Christodoulou}}\ and\ \bibinfo {author} {\bibfnamefont {C.}~\bibnamefont
  {Rovelli}},\ }\href {\doibase 10.1016/j.physletb.2019.03.015} {\bibfield
  {journal} {\bibinfo  {journal} {Physics Letters B}\ }\textbf {\bibinfo
  {volume} {792}},\ \bibinfo {pages} {64} (\bibinfo {year} {2019})}\BibitemShut
  {NoStop}%
\bibitem [{\citenamefont {Marletto}\ and\ \citenamefont
  {Vedral}(2020)}]{MarlettoVedral2020}%
  \BibitemOpen
  \bibfield  {author} {\bibinfo {author} {\bibfnamefont {C.}~\bibnamefont
  {Marletto}}\ and\ \bibinfo {author} {\bibfnamefont {V.}~\bibnamefont
  {Vedral}},\ }\href {\doibase 10.1103/PhysRevD.102.086012} {\bibfield
  {journal} {\bibinfo  {journal} {Phys. Rev. D}\ }\textbf {\bibinfo {volume}
  {102}},\ \bibinfo {pages} {086012} (\bibinfo {year} {2020})}\BibitemShut
  {NoStop}%
\bibitem [{\citenamefont {Bose}\ and\ \citenamefont {Morley}(2018)}]{Bose2018}%
  \BibitemOpen
  \bibfield  {author} {\bibinfo {author} {\bibfnamefont {S.}~\bibnamefont
  {Bose}}\ and\ \bibinfo {author} {\bibfnamefont {G.~W.}\ \bibnamefont
  {Morley}},\ }\href {https://arxiv.org/abs/1810.07045} {\bibfield  {journal}
  {\bibinfo  {journal} {arXiv:1810.07045}\ } (\bibinfo {year}
  {2018})}\BibitemShut {NoStop}%
\bibitem [{\citenamefont {Pedernales}\ \emph {et~al.}(2020)\citenamefont
  {Pedernales}, \citenamefont {Morley},\ and\ \citenamefont
  {Plenio}}]{Pedernales2020}%
  \BibitemOpen
  \bibfield  {author} {\bibinfo {author} {\bibfnamefont {J.~S.}\ \bibnamefont
  {Pedernales}}, \bibinfo {author} {\bibfnamefont {G.~W.}\ \bibnamefont
  {Morley}}, \ and\ \bibinfo {author} {\bibfnamefont {M.~B.}\ \bibnamefont
  {Plenio}},\ }\href {\doibase 10.1103/PhysRevLett.125.023602} {\bibfield
  {journal} {\bibinfo  {journal} {Phys. Rev. Lett.}\ }\textbf {\bibinfo
  {volume} {125}},\ \bibinfo {pages} {023602} (\bibinfo {year}
  {2020})}\BibitemShut {NoStop}%
\bibitem [{\citenamefont {Chevalier}\ \emph {et~al.}(2020)\citenamefont
  {Chevalier}, \citenamefont {Paige},\ and\ \citenamefont {Kim}}]{Kim2020}%
  \BibitemOpen
  \bibfield  {author} {\bibinfo {author} {\bibfnamefont {H.}~\bibnamefont
  {Chevalier}}, \bibinfo {author} {\bibfnamefont {A.~J.}\ \bibnamefont
  {Paige}}, \ and\ \bibinfo {author} {\bibfnamefont {M.~S.}\ \bibnamefont
  {Kim}},\ }\href {\doibase 10.1103/PhysRevA.102.022428} {\bibfield  {journal}
  {\bibinfo  {journal} {Phys. Rev. A}\ }\textbf {\bibinfo {volume} {102}},\
  \bibinfo {pages} {022428} (\bibinfo {year} {2020})}\BibitemShut {NoStop}%
\bibitem [{\citenamefont {Marshman}\ \emph {et~al.}(2020)\citenamefont
  {Marshman}, \citenamefont {Mazumdar},\ and\ \citenamefont
  {Bose}}]{Marshman2020}%
  \BibitemOpen
  \bibfield  {author} {\bibinfo {author} {\bibfnamefont {R.~J.}\ \bibnamefont
  {Marshman}}, \bibinfo {author} {\bibfnamefont {A.}~\bibnamefont {Mazumdar}},
  \ and\ \bibinfo {author} {\bibfnamefont {S.}~\bibnamefont {Bose}},\ }\href
  {\doibase 10.1103/PhysRevA.101.052110} {\bibfield  {journal} {\bibinfo
  {journal} {Phys. Rev. A}\ }\textbf {\bibinfo {volume} {101}},\ \bibinfo
  {pages} {052110} (\bibinfo {year} {2020})}\BibitemShut {NoStop}%
\bibitem [{\citenamefont {van~de Kamp}\ \emph {et~al.}(2020)\citenamefont
  {van~de Kamp}, \citenamefont {Marshman}, \citenamefont {Bose},\ and\
  \citenamefont {Mazumdar}}]{vandeKamp2020}%
  \BibitemOpen
  \bibfield  {author} {\bibinfo {author} {\bibfnamefont {T.~W.}\ \bibnamefont
  {van~de Kamp}}, \bibinfo {author} {\bibfnamefont {R.~J.}\ \bibnamefont
  {Marshman}}, \bibinfo {author} {\bibfnamefont {S.}~\bibnamefont {Bose}}, \
  and\ \bibinfo {author} {\bibfnamefont {A.}~\bibnamefont {Mazumdar}},\ }\href
  {\doibase 10.1103/PhysRevA.102.062807} {\bibfield  {journal} {\bibinfo
  {journal} {Phys. Rev. A}\ }\textbf {\bibinfo {volume} {102}},\ \bibinfo
  {pages} {062807} (\bibinfo {year} {2020})}\BibitemShut {NoStop}%
\bibitem [{\citenamefont {Page}\ and\ \citenamefont
  {Geilker}(1981)}]{Page1981}%
  \BibitemOpen
  \bibfield  {author} {\bibinfo {author} {\bibfnamefont {D.~N.}\ \bibnamefont
  {Page}}\ and\ \bibinfo {author} {\bibfnamefont {C.~D.}\ \bibnamefont
  {Geilker}},\ }\href {\doibase 10.1103/physrevlett.47.979} {\bibfield
  {journal} {\bibinfo  {journal} {Phys. Rev. Lett.}\ }\textbf {\bibinfo
  {volume} {47}},\ \bibinfo {pages} {979} (\bibinfo {year} {1981})}\BibitemShut
  {NoStop}%
\bibitem [{\citenamefont {Hawkins}(1982)}]{Hawkins1982}%
  \BibitemOpen
  \bibfield  {author} {\bibinfo {author} {\bibfnamefont {B.}~\bibnamefont
  {Hawkins}},\ }\href {\doibase 10.1103/physrevlett.48.520} {\bibfield
  {journal} {\bibinfo  {journal} {Phys. Rev. Lett.}\ }\textbf {\bibinfo
  {volume} {48}},\ \bibinfo {pages} {520} (\bibinfo {year} {1982})}\BibitemShut
  {NoStop}%
\bibitem [{\citenamefont {Ballentine}(1982)}]{Ballentine1982}%
  \BibitemOpen
  \bibfield  {author} {\bibinfo {author} {\bibfnamefont {L.~E.}\ \bibnamefont
  {Ballentine}},\ }\href {\doibase 10.1103/PhysRevLett.48.522} {\bibfield
  {journal} {\bibinfo  {journal} {Phys. Rev. Lett.}\ }\textbf {\bibinfo
  {volume} {48}},\ \bibinfo {pages} {522} (\bibinfo {year} {1982})}\BibitemShut
  {NoStop}%
\bibitem [{\citenamefont {Page}(1982)}]{Page1982}%
  \BibitemOpen
  \bibfield  {author} {\bibinfo {author} {\bibfnamefont {D.~N.}\ \bibnamefont
  {Page}},\ }\href {\doibase 10.1103/PhysRevLett.48.523} {\bibfield  {journal}
  {\bibinfo  {journal} {Phys. Rev. Lett.}\ }\textbf {\bibinfo {volume} {48}},\
  \bibinfo {pages} {523} (\bibinfo {year} {1982})}\BibitemShut {NoStop}%
\bibitem [{\citenamefont {Wolf}\ \emph {et~al.}(2003)\citenamefont {Wolf},
  \citenamefont {Eisert},\ and\ \citenamefont {Plenio}}]{WolfEisertPlenio2003}%
  \BibitemOpen
  \bibfield  {author} {\bibinfo {author} {\bibfnamefont {M.~M.}\ \bibnamefont
  {Wolf}}, \bibinfo {author} {\bibfnamefont {J.}~\bibnamefont {Eisert}}, \ and\
  \bibinfo {author} {\bibfnamefont {M.~B.}\ \bibnamefont {Plenio}},\ }\href
  {\doibase 10.1103/PhysRevLett.90.047904} {\bibfield  {journal} {\bibinfo
  {journal} {Phys. Rev. Lett.}\ }\textbf {\bibinfo {volume} {90}},\ \bibinfo
  {pages} {047904} (\bibinfo {year} {2003})}\BibitemShut {NoStop}%
\bibitem [{\citenamefont {Mancini}\ \emph {et~al.}(1998)\citenamefont
  {Mancini}, \citenamefont {Vitali},\ and\ \citenamefont
  {Tombesi}}]{Mancini1998}%
  \BibitemOpen
  \bibfield  {author} {\bibinfo {author} {\bibfnamefont {S.}~\bibnamefont
  {Mancini}}, \bibinfo {author} {\bibfnamefont {D.}~\bibnamefont {Vitali}}, \
  and\ \bibinfo {author} {\bibfnamefont {P.}~\bibnamefont {Tombesi}},\ }\href
  {\doibase 10.1103/PhysRevLett.80.688} {\bibfield  {journal} {\bibinfo
  {journal} {Phys. Rev. Lett.}\ }\textbf {\bibinfo {volume} {80}},\ \bibinfo
  {pages} {688} (\bibinfo {year} {1998})}\BibitemShut {NoStop}%
\bibitem [{\citenamefont {Cohadon}\ \emph {et~al.}(1999)\citenamefont
  {Cohadon}, \citenamefont {Heidmann},\ and\ \citenamefont
  {Pinard}}]{Cohadon1999}%
  \BibitemOpen
  \bibfield  {author} {\bibinfo {author} {\bibfnamefont {P.~F.}\ \bibnamefont
  {Cohadon}}, \bibinfo {author} {\bibfnamefont {A.}~\bibnamefont {Heidmann}}, \
  and\ \bibinfo {author} {\bibfnamefont {M.}~\bibnamefont {Pinard}},\ }\href
  {\doibase 10.1103/PhysRevLett.83.3174} {\bibfield  {journal} {\bibinfo
  {journal} {Phys. Rev. Lett.}\ }\textbf {\bibinfo {volume} {83}},\ \bibinfo
  {pages} {3174} (\bibinfo {year} {1999})}\BibitemShut {NoStop}%
\bibitem [{\citenamefont {Marquardt}\ \emph {et~al.}(2007)\citenamefont
  {Marquardt}, \citenamefont {Chen}, \citenamefont {Clerk},\ and\ \citenamefont
  {Girvin}}]{Marquardt2007}%
  \BibitemOpen
  \bibfield  {author} {\bibinfo {author} {\bibfnamefont {F.}~\bibnamefont
  {Marquardt}}, \bibinfo {author} {\bibfnamefont {J.~P.}\ \bibnamefont {Chen}},
  \bibinfo {author} {\bibfnamefont {A.~A.}\ \bibnamefont {Clerk}}, \ and\
  \bibinfo {author} {\bibfnamefont {S.~M.}\ \bibnamefont {Girvin}},\ }\href
  {\doibase 10.1103/PhysRevLett.99.093902} {\bibfield  {journal} {\bibinfo
  {journal} {Phys. Rev. Lett.}\ }\textbf {\bibinfo {volume} {99}},\ \bibinfo
  {pages} {093902} (\bibinfo {year} {2007})}\BibitemShut {NoStop}%
\bibitem [{\citenamefont {Wilson-Rae}\ \emph {et~al.}(2007)\citenamefont
  {Wilson-Rae}, \citenamefont {Nooshi}, \citenamefont {Zwerger},\ and\
  \citenamefont {Kippenberg}}]{Wilson-Rae2007}%
  \BibitemOpen
  \bibfield  {author} {\bibinfo {author} {\bibfnamefont {I.}~\bibnamefont
  {Wilson-Rae}}, \bibinfo {author} {\bibfnamefont {N.}~\bibnamefont {Nooshi}},
  \bibinfo {author} {\bibfnamefont {W.}~\bibnamefont {Zwerger}}, \ and\
  \bibinfo {author} {\bibfnamefont {T.~J.}\ \bibnamefont {Kippenberg}},\ }\href
  {\doibase 10.1103/PhysRevLett.99.093901} {\bibfield  {journal} {\bibinfo
  {journal} {Phys. Rev. Lett.}\ }\textbf {\bibinfo {volume} {99}},\ \bibinfo
  {pages} {093901} (\bibinfo {year} {2007})}\BibitemShut {NoStop}%
\bibitem [{\citenamefont {Aasi}\ and\ \citenamefont {\emph{et
  al.}}(2015)}]{ALIGO2015}%
  \BibitemOpen
  \bibfield  {author} {\bibinfo {author} {\bibfnamefont {J.}~\bibnamefont
  {Aasi}}\ and\ \bibinfo {author} {\bibnamefont {\emph{et al.}}},\ }\href
  {\doibase 10.1088/0264-9381/32/7/074001} {\bibfield  {journal} {\bibinfo
  {journal} {Classical and Quantum Gravity}\ }\textbf {\bibinfo {volume}
  {32}},\ \bibinfo {pages} {074001} (\bibinfo {year} {2015})}\BibitemShut
  {NoStop}%
\bibitem [{\citenamefont {Thorne}\ and\ \citenamefont
  {Winstein}(1999)}]{Thorne1999}%
  \BibitemOpen
  \bibfield  {author} {\bibinfo {author} {\bibfnamefont {K.~S.}\ \bibnamefont
  {Thorne}}\ and\ \bibinfo {author} {\bibfnamefont {C.~J.}\ \bibnamefont
  {Winstein}},\ }\href {https://doi.org/10.1103/physrevd.60.082001} {\bibfield
  {journal} {\bibinfo  {journal} {Physical Review D}\ }\textbf {\bibinfo
  {volume} {60}} (\bibinfo {year} {1999})}\BibitemShut {NoStop}%
\bibitem [{\citenamefont {Chen}(2013)}]{Chen2013}%
  \BibitemOpen
  \bibfield  {author} {\bibinfo {author} {\bibfnamefont {Y.}~\bibnamefont
  {Chen}},\ }\href {\doibase 10.1088/0953-4075/46/10/104001} {\bibfield
  {journal} {\bibinfo  {journal} {Journal of Physics B: Atomic, Molecular and
  Optical Physics}\ }\textbf {\bibinfo {volume} {46}},\ \bibinfo {pages}
  {104001} (\bibinfo {year} {2013})}\BibitemShut {NoStop}%
\bibitem [{\citenamefont {Caves}\ and\ \citenamefont
  {Schumaker}(1985)}]{Caves1985a}%
  \BibitemOpen
  \bibfield  {author} {\bibinfo {author} {\bibfnamefont {C.~M.}\ \bibnamefont
  {Caves}}\ and\ \bibinfo {author} {\bibfnamefont {B.~L.}\ \bibnamefont
  {Schumaker}},\ }\href
  {https://journals.aps.org/pra/abstract/10.1103/PhysRevA.31.3068} {\bibfield
  {journal} {\bibinfo  {journal} {Phys. Rev. A}\ }\textbf {\bibinfo {volume}
  {31}},\ \bibinfo {pages} {3068} (\bibinfo {year} {1985})}\BibitemShut
  {NoStop}%
\bibitem [{\citenamefont {Kimble}\ \emph {et~al.}(2001)\citenamefont {Kimble},
  \citenamefont {Levin}, \citenamefont {Matsko}, \citenamefont {Thorne},\ and\
  \citenamefont {Vyatchanin}}]{Kimble02}%
  \BibitemOpen
  \bibfield  {author} {\bibinfo {author} {\bibfnamefont {H.~J.}\ \bibnamefont
  {Kimble}}, \bibinfo {author} {\bibfnamefont {Y.}~\bibnamefont {Levin}},
  \bibinfo {author} {\bibfnamefont {A.~B.}\ \bibnamefont {Matsko}}, \bibinfo
  {author} {\bibfnamefont {K.~S.}\ \bibnamefont {Thorne}}, \ and\ \bibinfo
  {author} {\bibfnamefont {S.~P.}\ \bibnamefont {Vyatchanin}},\ }\href
  {http://journals.aps.org/prd/abstract/10.1103/PhysRevD.65.022002} {\bibfield
  {journal} {\bibinfo  {journal} {Phys. Rev. D}\ }\textbf {\bibinfo {volume}
  {65}},\ \bibinfo {pages} {022002} (\bibinfo {year} {2001})}\BibitemShut
  {NoStop}%
\bibitem [{\citenamefont {Purdy}\ \emph {et~al.}(2013)\citenamefont {Purdy},
  \citenamefont {Yu}, \citenamefont {Peterson}, \citenamefont {Kampel},\ and\
  \citenamefont {Regal}}]{Purdy2013b}%
  \BibitemOpen
  \bibfield  {author} {\bibinfo {author} {\bibfnamefont {T.~P.}\ \bibnamefont
  {Purdy}}, \bibinfo {author} {\bibfnamefont {P.~L.}\ \bibnamefont {Yu}},
  \bibinfo {author} {\bibfnamefont {R.~W.}\ \bibnamefont {Peterson}}, \bibinfo
  {author} {\bibfnamefont {N.~S.}\ \bibnamefont {Kampel}}, \ and\ \bibinfo
  {author} {\bibfnamefont {C.~A.}\ \bibnamefont {Regal}},\ }\href
  {http://journals.aps.org/prx/abstract/10.1103/PhysRevX.3.031012} {\bibfield
  {journal} {\bibinfo  {journal} {Phys. Rev. X}\ }\textbf {\bibinfo {volume}
  {3}},\ \bibinfo {pages} {031012} (\bibinfo {year} {2013})}\BibitemShut
  {NoStop}%
\bibitem [{\citenamefont {Buchmann}\ \emph {et~al.}(2016)\citenamefont
  {Buchmann}, \citenamefont {Schreppler}, \citenamefont {Kohler}, \citenamefont
  {Spethmann},\ and\ \citenamefont {Stamper-Kurn}}]{Buchmann2016}%
  \BibitemOpen
  \bibfield  {author} {\bibinfo {author} {\bibfnamefont {L.~F.}\ \bibnamefont
  {Buchmann}}, \bibinfo {author} {\bibfnamefont {S.}~\bibnamefont
  {Schreppler}}, \bibinfo {author} {\bibfnamefont {J.}~\bibnamefont {Kohler}},
  \bibinfo {author} {\bibfnamefont {N.}~\bibnamefont {Spethmann}}, \ and\
  \bibinfo {author} {\bibfnamefont {D.~M.}\ \bibnamefont {Stamper-Kurn}},\
  }\href {\doibase 10.1103/PhysRevLett.117.030801} {\bibfield  {journal}
  {\bibinfo  {journal} {Phys. Rev. Lett.}\ }\textbf {\bibinfo {volume} {117}},\
  \bibinfo {pages} {030801} (\bibinfo {year} {2016})}\BibitemShut {NoStop}%
\bibitem [{\citenamefont {Yang}\ \emph {et~al.}(2013)\citenamefont {Yang},
  \citenamefont {Miao}, \citenamefont {Lee}, \citenamefont {Helou},\ and\
  \citenamefont {Chen}}]{Yang2013a}%
  \BibitemOpen
  \bibfield  {author} {\bibinfo {author} {\bibfnamefont {H.}~\bibnamefont
  {Yang}}, \bibinfo {author} {\bibfnamefont {H.}~\bibnamefont {Miao}}, \bibinfo
  {author} {\bibfnamefont {D.}~\bibnamefont {Lee}}, \bibinfo {author}
  {\bibfnamefont {B.}~\bibnamefont {Helou}}, \ and\ \bibinfo {author}
  {\bibfnamefont {Y.}~\bibnamefont {Chen}},\ }\href
  {https://journals.aps.org/prl/abstract/10.1103/PhysRevLett.110.170401}
  {\bibfield  {journal} {\bibinfo  {journal} {Phys. Rev. Lett.}\ }\textbf
  {\bibinfo {volume} {110}},\ \bibinfo {pages} {170401} (\bibinfo {year}
  {2013})}\BibitemShut {NoStop}%
\bibitem [{\citenamefont {Adesso}\ and\ \citenamefont
  {Illuminati}(2007)}]{Adesso2007}%
  \BibitemOpen
  \bibfield  {author} {\bibinfo {author} {\bibfnamefont {G.}~\bibnamefont
  {Adesso}}\ and\ \bibinfo {author} {\bibfnamefont {F.}~\bibnamefont
  {Illuminati}},\ }\href {\doibase 10.1088/1751-8113/40/28/s01} {\bibfield
  {journal} {\bibinfo  {journal} {Journal of Physics A: Mathematical and
  Theoretical}\ }\textbf {\bibinfo {volume} {40}},\ \bibinfo {pages} {7821}
  (\bibinfo {year} {2007})}\BibitemShut {NoStop}%
\end{thebibliography}%


\end{document}